\documentclass[12pt,epsf,amssymb]{article} \textheight =23 truecm
\def\lsim{\raise0.3ex\hbox{$<$\kern-0.75em\raise-1.1ex\hbox{$\sim$}}}
\def\gsim{\raise0.3ex\hbox{$>$\kern-0.75em\raise-1.1ex\hbox{$\sim$}}}
\def\noi{\noindent} \def\nn{\nonumber} \def\bea{\begin{eqnarray}}
\def\eea{\end{eqnarray}} \def\beq{\begin{equation}}
\def\eeq{\end{equation}} 
\def\beeq{\begin{eqnarray}} \def\eeeq{\end{eqnarray}} \def\R{ {\rm R
\kern -.31cm I \kern .15cm}} \def\C{ {\rm C \kern -.15cm \vrule
width.5pt \kern .12cm}} \def\Z{ {\rm Z \kern -.27cm \angle \kern
.02cm}} \def\N{ {\rm N \kern -.26cm \vrule width.4pt \kern .10cm}}
\def\1{{\rm 1\mskip-4.5mu l} }

\usepackage{graphics} \usepackage{graphicx} \usepackage{epsfig}

\begin{document} \begin{center} {\large \bf The tensor force in HQET and the semileptonic 
\vskip 3 truemm
{\large \bf  $\overline{\bf B}$ decay} to excited vector mesons ${\bf D\left ( \textstyle{{3 \over 2}^-,
1^-}\right )}$}

\vskip 1 truecm {\bf F. Jugeau, A. Le Yaouanc, L. Oliver and J.-C.
Raynal}\\

{\it Laboratoire de Physique Th\'eorique}\footnote{Unit\'e Mixte de
Recherche UMR 8627 - CNRS }\\    {\it Universit\'e de Paris XI,
B\^atiment 210, 91405 Orsay Cedex, France} \end{center}

\vskip 1 truecm

\begin{abstract} 
We extend the formalism of Leibovich, Ligeti, Stewart and Wise  in the $1/m_Q$ expansion of Heavy Quark
Effective Theory for the
$\overline{B}$ semileptonic decays into excited $D\left ( \textstyle{{3 \over 2}^+}\right )$ mesons to the opposite parity states $D\left ( \textstyle{{3 \over 2}^-}\right )$. For $D\left ( \textstyle{{3 \over 2}^+}\right )$ the $1/m_Q$ current perturbation dominates over the leading term at zero recoil, while for $D\left ( \textstyle{{3 \over 2}^-}\right )$ the $1/m_Q$ perturbation due to ${\cal L}_{mag}$ dominates also at zero recoil. We show that the corresponding $1/m_Q$
magnetic coupling is proportional to the mixing between the states $D\left ( \textstyle{{3 \over 2}^-,
1^-}\right )$ and $D\left ( \textstyle{{1 \over 2}^-, 1^-}\right )$ induced by the tensor force.
We point out some subtleties that appear in this respect in HQET.
\end{abstract}

\vskip 2 truecm

\noi LPT Orsay 05-44 \par \noi July 2005
\par \vskip 1 truecm

\noindent e-mails : frederic.jugeau@th.u-psud.fr,
leyaouan@th.u-psud.fr, oliver@th.u-psud.fr 

\newpage \pagestyle{plain}
In this note we deal with the relation between the tensor force and weak transitions at zero recoil in Heavy Quark
Effective Theory (HQET), and some subtleties related to this
question.\par

In the quark model, the tensor force between two quarks $Q$ and
$\overline{q}$ of unequal masses is given by the expression \cite{1r}
\beq
\label{1e}
H_{tensor}^{Qq} = {1 \over m_Q m_q} U(r_{Qq}) \left [ {3 ({\bf S}_Q \cdot {\bf r}_{Qq}) ({\bf S}_q \cdot {\bf r}_{Qq}) \over r_{Qq}^2} - ({\bf S}_Q \cdot {\bf S}_q)\right ]
 \eeq
 
\noi where, for one-gluon exchange, $U(r_{Qq})$ is positive and
proportional to ${\alpha_S \over r_{Qq}^3}$. \par

 Expression (\ref{1e}) shows that in the case of heavy-light mesons ($m_Q
\gg m_q)$, $H_{tensor}^{Qq}$ is proportional to $1/m_Q$. Therefore, the
tensor force should appear in HQET at the first order in the $1/m_Q$
expansion.\par

In mesons, this force is responsible for mixing between the vector states $D\left
( \textstyle{{3 \over 2}^-, 1^-}\right )$ and $D\left (
\textstyle{{1 \over 2}^-, 1^-}\right )$. In the quark model, a
straighforward calculation using (\ref{1e}) gives
\beq
\label{2e}
<D\left ( \textstyle{{3 \over 2}^-, 1^-}\right )|H_{tensor}^{Qq}|D\left ( \textstyle{{1 \over 2}^-, 1^-}\right )>\ \sim \ {<\psi_1|U(r_{Qq})|\psi_0 > \over m_Q m_q}
\eeq

\noi where $\psi_L(r_{Qq})$ ($L = 0, 1$) are the radial wave functions
of the ground state and of the first orbitally excited state.\par

In HQET, the mixing between $D\left ( \textstyle{{3 \over 2}^-,
1^-}\right )$ and $D\left ( \textstyle{{1 \over 2}^-, 1^-}\right
)$, equivalent to (\ref{2e}), will be given by the matrix element
\beq
\label{3e}
<D\left ( \textstyle{{3 \over 2}^-, 1^-}\right )(v, \varepsilon ) |{\cal L}_{mag, v}^{(c)}(0) | D\left ( \textstyle{{1 \over 2}^-, 1^-}\right )(v, \varepsilon )> 
\eeq

\noi with
\bea
\label{4e}
&&{\cal L}_{mag, v}^{(Q)} = {1 \over 2m_Q} O_{mag, v}^{(Q)} \nn \\
&&O_{mag, v}^{(Q)} = {g_s \over 2} \ \overline{h}_v^{(Q)} \sigma_{\alpha\beta}G^{\alpha\beta}h_v^{(Q)} 
\eea

The ${3 \over 2}^-$ and the
${1 \over 2}^-$ fields of spin 1 are given by \cite{2r} \cite{3r}
\bea
\label{6e}
&&H_v\left ( \textstyle{{1 \over 2}^-, 1^-}\right ) = P_+ {/\hskip - 2 truemm \varepsilon}_v\nn \\
&&F_v^{\sigma} \left ( \textstyle{{3 \over 2}^-, 1^-}\right ) =  \sqrt{{3\over 2}}P_+ \varepsilon_v^{\rho} \left [ g_{\rho}^{\sigma} - {1 \over 3} \gamma_{\rho} (\gamma^{\sigma} + v^{\sigma})\right ] 
\eea

\noi where $P_+ = \displaystyle{{1  +  {/ \hskip - 2 truemm v}} \over 2}$ and the last
expression follows from 
\beq
\label{6new}
F_v^{\sigma} \left ( \textstyle{{3 \over 2}^+, 1^+}\right ) = - \sqrt{{3\over 2}}P_+ \varepsilon_v^{\rho} \gamma_5 \left [ g_{\rho}^{\sigma} - {1 \over 3} \gamma_{\rho} (\gamma^{\sigma} - v^{\sigma})\right ]
\eeq 

\noi  multiplying by $(-\gamma_5)$ on
the right.\par

The mixing will then be given by the matrix element
\beq
\label{7new}
{<D\left ( \textstyle{{3 \over 2}^{-}, 1^{-}}\right )(v, \varepsilon)|{\cal L}_{mag, v}^{(c)}(0) |D\left ( \textstyle{{1 \over 2}^{-}, 1^{-}}\right )(v, \varepsilon)>\over \sqrt{m_{D_{3/2}} \ m_{D_{1/2}}}} = {1 \over 2m_c} Tr \left [ M_{\sigma\alpha\beta}^{(c)} \overline{F}_v^{\sigma} i \sigma^{\alpha\beta}H_v \right ]
\eeq

\noi where
\beq
\label{8new}
\overline{F}_v^{\sigma} = \gamma^0 F_v^{\sigma^+} \gamma^0 = \sqrt{{3 \over 2}} \left [ g_{\rho}^{\sigma} - {1 \over 3} \left ( \gamma^{\sigma} + v^{\sigma}\right ) \gamma_{\rho}\right ] \varepsilon_v^{*\rho} P_+ \ .
\eeq

Since $\gamma_{\sigma} \overline{F}_v^{\sigma} = 0$, the Dirac structure of $M_{\sigma\alpha\beta}^{(c)}$ could contain terms of the form $v_{\sigma}\gamma_{\alpha} \gamma_{\beta}$, $v_{\sigma}v_{\alpha} \gamma_{\beta}$, $g_{\sigma\alpha} v_{\beta}$ and $g_{\sigma\alpha} \gamma_{\beta}$. However, since the matrix element (\ref{7new}) is at zero recoil, one has  
\beq
\label{9new}
v_{\sigma} \overline{F}_v^{\sigma}  = v_{\beta} P_+ i \sigma^{\alpha\beta} P_+ = 0
\eeq

\noi and the only surviving term has the form $g_{\sigma\alpha} \gamma_{\beta}$. Therefore
\beq
\label{10new}
M_{\sigma\alpha\beta}^{(c)} = \mu  \ g_{\sigma\alpha} \gamma_{\beta}
\eeq

\noi and the mixing matrix element (\ref{7new}) is proportional to the coupling $\mu$,
\beq
\label{11new}
{<D\left ( \textstyle{{3 \over 2}^{-}, 1^{-}}\right )(v, \varepsilon)|{\cal L}_{mag, v}^{(c)}(0) |D\left ( \textstyle{{1 \over 2}^{-}, 1^{-}}\right )(v, \varepsilon)>\over  \sqrt{m_{D_{3/2}} \ m_{D_{1/2}}}} = - 4 \sqrt{{2 \over 3}} \ {\mu \over 2m_c}
\eeq

To see how in HQET the transition $B \to D\left ( \textstyle{{3 \over 2}^{-}, 1^{-}}\right )$ at first order in
$1/m_Q$ at zero recoil is related to this mixing, let us use the formalism of Leibovich,
Ligeti, Stewart and Wise \cite{2r}, that was applied to ${1\over 2}^-
\to {3 \over 2}^+$ transitions. Leibovich et al have considered already the $B \to D\left ( \textstyle{{3 \over 2}^{-}, 1^{-}}\right )$ transition at zero recoil (Section IV of \cite{2r}). However, for our purpose, it will be instructive to use the general formalism, and consider here the matrix elements
{\it at non-zero recoil} to compare both cases ${1\over 2}^- \to
{3\over 2}^+$, ${1\over 2}^- \to {3 \over 2}^-$. At the end of the calculation we will take the zero recoil limit for the $B \to D\left ( \textstyle{{3 \over 2}^{-}, 1^{-}}\right )$ transition. \par

The study of the semileptonic decay $\overline{B} \to D\left ( \textstyle{{3 \over 2}^{-}, 1^{-}}\right )\ell \overline{\nu}_{\ell}$ is not only of academic interest, since such orbitally excited state is expected at a mass $\lsim\ 2.8$~GeV. However, we expect this state to be wide, since it can decay, among other modes, by $S$-wave into $D\left ( \textstyle{{3 \over 2}^{+}, 1^{+}}\right ) + \pi$.\par

First, we must notice that at non-zero recoil, the $1/m_Q$ perturbations to the matrix elements
\beq
\label{12new}
<D\left ( \textstyle{{3 \over 2}^{-}, 1^{-}}\right )(v') | [ \overline{h}_{v'}^{(c)} \Gamma h_v^{(b)}] (0) |B(v)>
\eeq

\noi are of three types : current perturbations and perturbations of the lagrangian ${\cal L}_{kin}$ and ${\cal L}_{mag}$. The leading order matrix element (\ref{12new}) vanishes at zero recoil \cite{2r}, and the same happens for the ${\cal L}_{kin}$ perturbation, that behaves in powers of ($w-1$) as the leading term.\par

Concerning the current perturbation matrix element, it also vanishes at zero recoil, as pointed out in ref. \cite{2r} (Section IV). This follows from the relation ${/\hskip - 2 truemm v} F_v^{\sigma} = F_v^{\sigma} = F_v^{\sigma}{/\hskip - 2 truemm v}$, that can be read from (\ref{6e}). This is at odds with the current perturbation matrix element for $B(v) \to  D\left ( \textstyle{{3 \over 2}^{+}, 1^{+}}\right )(v')$, that, in general, does not vanish at zero recoil \cite{2r}.\par

Therefore, we will only consider matrix elements of the ${\cal L}_{mag}$ perturbation, but study in parallel the $B(v) \to D\left ( \textstyle{{3 \over 2}^{\pm}, 1^{\pm }}\right )(v')$ transitions, to grasp the difference between both cases. As we will see below, due ${\cal L}_{mag}$, the transition $B(v) \to  D\left ( \textstyle{{3 \over 2}^{-}, 1^{-}}\right )(v')$, unlike $B(v) \to  D\left ( \textstyle{{3 \over 2}^{+}, 1^{+}}\right )(v')$, does not vanish at zero recoil.

Considering an arbitrary current
$\overline{c}\Gamma b$, the relevant matrix elements are
\bea
\label{5e}
&&{1 \over \sqrt{m_{D_{3/2}} \ m_B}}\ <D\left ( \textstyle{{3 \over 2}^{\pm}, 1^{\pm}}\right )(v') |i \int dxT \left \{ {\cal L}_{mag, v'}^{(c)}(x)  [  \overline{h}_{v'}^{(c)}\Gamma h_v^{(b)}] (0)\right \}\nn \\
&&+ \ i \int dxT \left \{ {\cal L}_{mag, v}^{(b)}(x)  [  \overline{h}_{v'}^{(c)}\Gamma h_v^{(b)} ] (0)\right \} |B (v) > \nn \\
&&= {1 \over 2m_c} Tr \left [ R_{\sigma \alpha \beta}^{(\pm)(c)} \overline{F}_{v'}^{(\pm ) \sigma} i \sigma^{\alpha\beta}P'_+\Gamma H_v\right ] + {1 \over 2m_b} Tr \left [ R_{\sigma \alpha \beta}^{(\pm)(b)} \overline{F}_{v'}^{(\pm ) \sigma} \Gamma P_+i \sigma^{\alpha\beta}H_v\right ] 
\eea 

\noi where the superindex $\pm$  in $F_{v'}^{(\pm)\sigma}$ indicates the
parity of the state $D\left ( \textstyle{{3 \over 2}^{\pm},
1^{\pm}}\right )$, $H_v$ corresponds to the pseudoscalar state~:
\beq
\label{14new}
H_v\left ( \textstyle{{1 \over 2}^{-}, 0^{-}}\right ) = P_+(-\gamma_5)
\eeq 

\noi and $m_{D_{3/2}}$ is the mass of either the ${3 \over 2}^+$ or the ${3 \over 2}^-$ meson. \par

Using the conditions
\beq
\label{7e}
F_v^{\sigma}v_{\sigma} = F_v^{\sigma}\gamma_{\sigma} = 0
\eeq

\noi and the antisymmetry of $i\sigma^{\alpha\beta}$, that implies $P_+
v_{\alpha} i\sigma^{\alpha\beta}P_+ = P'_+
v'_{\alpha}i\sigma^{\alpha\beta}P'_+ = 0$, the parametrizations for $R_{\sigma\alpha\beta}^{(\pm ) (c)}$, $R_{\sigma\alpha\beta}^{(\pm ) (b)}$ follow~:
\bea
\label{8e}
&&R^{(\pm)(c)}_{\sigma\alpha\beta} = \eta_1^{(\pm )(c)} v_{\sigma} \gamma_{\alpha} \gamma_{\beta} + \eta_2^{(\pm )(c)} v_{\sigma} v_{\alpha} \gamma_{\beta} + \eta_3^{(\pm )(c)} g_{\sigma\alpha} v_{\beta} + \eta_4^{(\pm )(c)} g_{\sigma\alpha} \gamma_{\beta}\nn \\
&&R^{(\pm)(b)}_{\sigma\alpha\beta} = \eta_1^{(\pm )(b)} v_{\sigma} \gamma_{\alpha} \gamma_{\beta} + \eta_2^{(\pm )(b)} v_{\sigma} v'_{\alpha} \gamma_{\beta} + \eta_3^{(\pm )(b)} g_{\sigma\alpha} v'_{\beta} + \eta_4^{(\pm )(b)} g_{\sigma\alpha} \gamma_{\beta}
\eea

\noi where the $\eta$'s depend on $w$. Other tensor structures give
terms that, under the trace, are linearly dependent on these
terms.\par

Owing to our remarks on the mixing (\ref{10new}) we have kept on purpose the term that has the tensor structure
$g_{\sigma\alpha}\gamma_{\beta}$. As pointed out by Leibovich et al.,
this term is not independent from the others for ${1 \over 2}^- \to {3 \over 2}^+$ transitions. They correctly choose,
for ${1\over 2}^- \to {3 \over 2}^+$, the basis (omitting the $(+)$ superindex)
\bea
\label{9e}
&&R_{\sigma\alpha\beta}^{(c)} = \eta _1^{(c)} \ v_{\sigma} \gamma_{\alpha} \gamma_{\beta} + \eta ^{(c)}_2 \ v_{\sigma} v_{\alpha} \gamma_{\beta} + \eta^{(c)}_3 \ g_{\sigma\alpha} v_{\beta}\nn \\
&&R_{\sigma\alpha\beta}^{(b)} = \eta_1^{(b)} \ v_{\sigma} \gamma_{\alpha} \gamma_{\beta} + \eta_2^{(b)} \ v_{\sigma} v'_{\alpha} \gamma_{\beta} + \eta_3^{(b)} \ g_{\sigma\alpha} v'_{\beta} \ .
\eea

However, this is not the natural basis for ${1\over 2}^- \to {3 \over
2}^-$ transitions. Indeed, from (\ref{5e})-(\ref{8e}) one gets, for three terms of $R_{\sigma\alpha\beta}^{(\pm)(c)}$ in
(\ref{8e}) the following trace identity, respectively for the
transitions ${1 \over 2}^- \to {3 \over 2}^+$, ${1 \over 2}^- \to {3
\over 2}^-$,
\beq
\label{10e}
Tr \left \{ \left [ v_{\sigma}\gamma_{\alpha}\gamma_{\beta} + 2 g_{\sigma\alpha} v_{\beta} + 2(1 \pm w) g_{\sigma \alpha} \gamma_{\beta} \right ] \overline{F}_{v'}^{(\pm )\sigma}i\sigma^{\alpha\beta}P'_+ \Gamma H_v \right \} = 0 \ .
\eeq

\noi While the
basis (\ref{9e}) is suitable for ${1 \over 2}^- \to {3 \over 2}^+$
transitions, this is not the case for ${1\over 2}^- \to {3 \over 2}^-$,
because $(1-w) g_{\sigma\alpha}\gamma_{\beta}$ {\it vanishes at zero
recoil}.\par

The decay matrix elements of $B$ mesons that are related to the
mixing between $D\left ( \textstyle{{3 \over 2}^{-}, 1^-}\right
)$ and $D\left ( \textstyle{{1 \over 2}^{-}, 1^-}\right )$, are the matrix element at zero recoil through the axial current~:
$${<D\left ( \textstyle{{3 \over 2}^{-}, 1^-}\right )(v, \varepsilon ) |i \int dxT \left \{ {\cal L}_{mag, v}^{(c)}(x)  [  \overline{h}_{v}^{(c)}\gamma_{\mu}\gamma_5 h_v^{(b)} ] (0)\right \}| B(v))>\over \sqrt{m_{D_{3/2}}\ m_B}}$$
\beq
= {1 \over 2m_c} Tr \left [ R_{\sigma \alpha \beta}^{(-)(c)} \overline{F}_{v}^{(-) \sigma} i \sigma^{\alpha\beta}P_+\gamma_{\mu}\gamma_5 H_v\right ] 
\label{11e}
\eeq 

\noi where we have used (\ref{5e}) and (\ref{8e}) and the contribution of $R_{\sigma\alpha\beta}^{(-)(b)}$ vanishes at zero recoil. \par

Using the decomposition (\ref{8e}) at zero recoil, we see that the
terms $v_{\sigma}\gamma_{\alpha}\gamma_{\beta}$, $v_{\sigma}v_{\alpha}\gamma_{\beta}$, $g_{\sigma\alpha}v_{\beta}$ do not contribute because of the relations (\ref{9new}), and  we are only left with the term $\eta_4^{(-)(c)}(1)g_{\sigma\alpha}\gamma_{\beta}$. One finds, after some
Dirac algebra,
\beq
\label{12e}
{<D\left ( \textstyle{{3 \over 2}^{-}, 1^-}\right )(v, \varepsilon) |i \int dxT \left \{ {\cal L}_{mag, v}^{(c)}(x)  [  \overline{h}_{v}^{(c)}\gamma_{\mu}\gamma_5 h_v^{(b)}] (0)\right \}| B(v)>\over \sqrt{m_{D_{3/2}}\ m_B}}
= 2 \sqrt{{2 \over 3}}  {\eta_4^{(-)(c)}(1) \over 2m_c} \varepsilon_{\mu}^*\ .
\eeq 

\noi On the other hand, one can insert intermediate states in the $T$-product and obtain
$${<D\left ( \textstyle{{3 \over 2}^{-}, 1^-}\right )(v, \varepsilon) |i \int dxT \left \{ {\cal L}_{mag, v}^{(c)}(x)  [  \overline{h}_{v}^{(c)}\gamma_{\mu}\gamma_5 h_v^{(b)}] (0)\right \}| B(v)>\over \sqrt{m_{D_{3/2}}\ m_B}}$$
\beq
\label{13e}
=  \varepsilon_{\mu}^*  {1 \over \Delta E} { <D\left ( \textstyle{{3 \over 2}^{-}, 1^-}\right )(v, \varepsilon)|{\cal L}_{mag, v}^{(c)}(0)|D\left ( \textstyle{{1 \over 2}^{-}, 1^-}\right )(v, \varepsilon)> \over \sqrt{m_{D_{3/2}} m_{D_{1/2}}}}
\eeq 

\noi where $\Delta E$ is the level spacing $\Delta E = m_{D_{3/2}} - m_{D_{1/2}}$. Only the $(n=0)$ ground state $D\left ( \textstyle{{1 \over 2}^{-}, 1^-}\right )$ contributes to the sum because the matrix element is at zero recoil and one has $\xi (1) = 1$, $\xi^{(n)}(1) = 0$ ($n \not= 0$). The factor in front of the r.h.s. of (\ref{13e}) comes from the calculation of the trace 
\beq
\label{22new}
 Tr \left [ {/\hskip - 2 truemm  \varepsilon }^* P_+  \gamma_{\mu} \gamma_5 P_+(- \gamma_5) \right ] = - 2  \varepsilon_{\mu}^*
\eeq

\noi Therefore, comparing (\ref{12e}) and (\ref{13e}) one finds
\beq
\label{23new}
2\sqrt{{2 \over 3}} {\eta_4^{(-)} \over 2m_c} =  {1 \over \Delta E} { <D\left ( \textstyle{{3 \over 2}^{-}, 1^-}\right )(v, \varepsilon)|{\cal L}_{mag, v}^{(c)}(0)|D\left ( \textstyle{{1 \over 2}^{-}, 1^-}\right )(v, \varepsilon)> \over \sqrt{m_{D_{3/2}} m_{D_{1/2}}}}
\eeq  

In conclusion, we have shown that  in HQET the transition between $B$ and $D\left ( \textstyle{{3
\over 2}^{-}, 1^-}\right )$ mesons  through the axial current at zero recoil is proportional to
the mixing between the states $D\left ( \textstyle{{3 \over 2}^{-}, 1^-}\right )$ and $D\left ( \textstyle{{1\over 2}^{-}, 1^-}\right )$ due to the tensor
force induced by ${\cal L}_{mag}$.

\section*{Acknowledgement}

We are indebted to the EC contract HPRN-CT-2002-00311 (EURIDICE).

\end{document}